\documentclass[11pt,twoside]{article}
\usepackage{APN5proc}

\resetcounters

\begin{document}

\title{UV Spectroscopy of the Central Star of the Planetary Nebula A\,43}
\author{Ellen Ringat, Felix Friederich, Thomas Rauch, and Klaus Werner}{\small \affil{Institute for Astronomy and Astrophysics, Kepler Center for Astro and Particle Physics, Eberhard Karls University, Sand 1, 72076 T\"ubingen, Germany}}
\author{Jeffrey W. Kruk}{\small \affil{NASA, Goddard Space Flight Center, Greenbelt, MD 20771, USA}
}

\begin{abstract}
About 25\% of all post-AGB stars are hydrogen-deficient, e.g. the PG\,1159 stars with a typical abundance pattern He:C:O = 33:50:17 (by mass). Only four of about 40 known PG\,1159 stars exhibit H in their spectra. The exciting star of the planetary nebula A\,43 is one of these so-called hybrid PG\,1159 stars. We present preliminary results of an on-going spectral analysis by means of NLTE model-atmosphere techniques based on UV spectra obtained with \emph{FUSE}, \emph{HST}/\emph{GHRS}, and \emph{IUE} as well as on optical observations. \\
\noindent{\bf Keywords.}\hspace{10pt}ISM: planetary nebulae: individual: A\,43 $-$ Stars: abundances $-$ Stars: atmospheres $-$ Stars: evolution $-$ Stars: individual: WD\,1721+106 $-$ Stars: AGB and post-AGB
\end{abstract}

\section{Introduction}
PG\,1159 stars are hydrogen-deficient post-AGB stars. They have temperatures between 75\,000\,K and 200\,000\,K and their surface gravities $\log g$ range between 5.5 and 8.0. They experienced a Final Thermal Pulse (FTP), that mixed the envelope and the intershell (Fig.\,1), and became a born-again star \citep{WernerHerwig}. Depending on the occurrence of this FTP a maximum remaining hydrogen content of about 20\% (by mass) is possible. This maximum value is predicted for the AGB Final Thermal Pulse (AFTP) scenario, where the FTP happens at the end of the AGB phase and the masses of the mixing shells are nearly the same. When the FTP happens later, it is called Late Thermal Pulse (LTP) and the hydrogen content of the star is about 1\% due to the smaller envelope mass at the time the FTP happens. An even later FTP (a Very Late Thermal Pulse occurs on the white dwarf cooling sequence) causes a complete burning of the hydrogen. Most PG\,1159 stars experienced a LTP or VLTP. Only four stars show hydrogen lines in their spectra which is a hint for an AFTP. Three of these so-called hybrid PG\,1159 stars (namely the central stars of Sh\,2$-$68, A\,43, and NGC\,7094) are surrounded by a planetary nebula. One of these hybrid PG\,1159 stars, A\,43, is used as an example to introduce the spectral analysis technique in Sect. 3. Then our preliminary results are discussed in Sect. 4. The opportunity to access \emph{TMAP} or already calculated \emph{TMAP} spectra via the VO service \textit{TheoSSA} is described in Sect. 5. 

\begin{minipage}{8cm}
\setlength{\textwidth}{8cm}
\setlength{\textheight}{5cm}
\hspace{-0.5cm}
\plotone[width=0.95\textwidth]{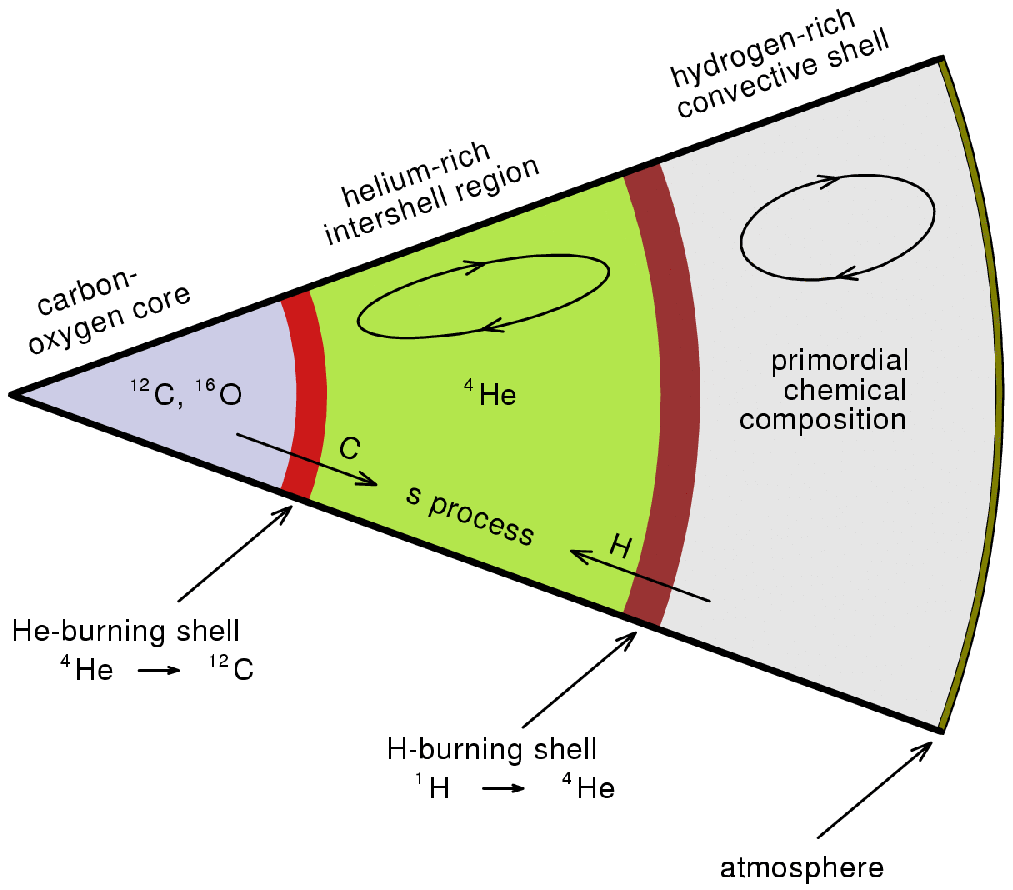}
\end{minipage}
\hbox{}\hspace{5mm}
\begin{minipage}{3.75cm}
\vspace{2.5cm}
\setlength{\textwidth}{3.75cm}
\setlength{\textheight}{3.75cm}
Figure 1. \,Structure of a post-AGB star. The intershell and the hydrogen-rich shell are mixed by a FTP.
\end{minipage}
\vspace{-1.6cm}

\section{Observational data}
To perform a precise analysis, observed spectra with a high S/N ratio and resolution are required. Moreover, these spectra should cover a wide wavelength range. Because of the high temperature of PG\,1159 stars, their maximum flux is located in the UV. A lot of strategic lines is located in this wavelength range, and thus, these spectra are very important for our analysis. To determine the surface gravity, optical spectra are advantageous. For our analysis, we retrieved \emph{FUSE}, \emph{GHRS}, and \emph{IUE} spectra from MAST (standard pipeline reduction, Table\,\ref{Observations}). The spectra for the optical wavelength range were obtained with the 3.5\,m telescope equipped with the TWIN spectrograph of the German-Spanish Astronomical Center on Calar Alto on April 13-14, 2001. They were reduced with the Image Reduction and Analysis Facility (IRAF, http://iraf.noao.edu/). To determine the interstellar extinction we used the UV spectra as well as brightnesses in the optical and infrared wavelength range and considered the Fitzpatrick law \citeyearpar{Fitzpatrick}. The resulting reddening is $E_{\mathrm{B}-\mathrm{V}}=0.265\pm0.035$.

\begin{table}[!ht]
\caption{Log of our UV observations.}
\label{Observations}
\smallskip
\begin{center}
{\scriptsize
\begin{tabular}{lrrrr}
\tableline
\noalign{\smallskip}
Instrument& Obs ID&Obs Start Time (UT)&Aperture&Exp. Time (sec)\\
\noalign{\smallskip}
\tableline
\noalign{\smallskip}
\emph{FUSE}&B0520201000&2001-07-29 20:41:47&LWRS&11438\\
\emph{FUSE}&B0520202000&2001-08-03 22:18:20&LWRS&9528\\
\emph{GHRS}&Z3GW0304T&1996-09-08 07:00:34  	&2.0&4243\\
\emph{IUE}&LWR08735&1980-09-06 21:45:21&LARGE&3600\\
\emph{IUE}&SWP10245&1980-09-28 21:50:02&LARGE&5100\\
\noalign{\smallskip}
\tableline
\end{tabular}
}
\end{center}
\end{table}

\section{Spectral analysis}
\par
The spectral analysis was performed with the T\"ubingen NLTE Model-Atmosphere Package (\emph{TMAP}) which was developed over the last 25 years. It uses an Accelerated Lambda Iteration (ALI, see e.g. \citealt{WernerDreizler,Werner2003,RauchDeetjen}). Hydrostatic and radiative equilibrium and plane-parallel geometry are assumed. About 1000 atomic levels can be considered as NLTE levels and thousands of individual lines can be calculated.  
\par
In the UV range numerous interstellar lines overlay the photospheric lines. The program OWENS was used to calculate the interstellar line-absorption spectrum. With OWENS various interstellar clouds, that consider different temperatures, radial and turbulent velocities, chemical compositions, and column densities can be modeled. 
\par
We used the parameters determined by Ziegler (priv. comm.) for the so-called spectroscopic twin of A\,43, namely NGC\,7094, as start values for our analysis ($T_\mathrm{eff}=105\,$kK, $\log g=5.4$). First, we checked these values with H+He models. Then we calculated more detailed model atmospheres for the newly determined values and included the elements H-Ni. After fitting the effective temperature and surface gravity, the elemental abundances are fine-tuned. In the UV range this is done in an iterative process with \emph{TMAP} and OWENS until the interstellar as well as the photospheric lines reproduce the observed lines (Fig.\,\ref{FUSE}). 
\setcounter{figure}{1}
\begin{figure}
\plotone[width=0.95\textwidth]{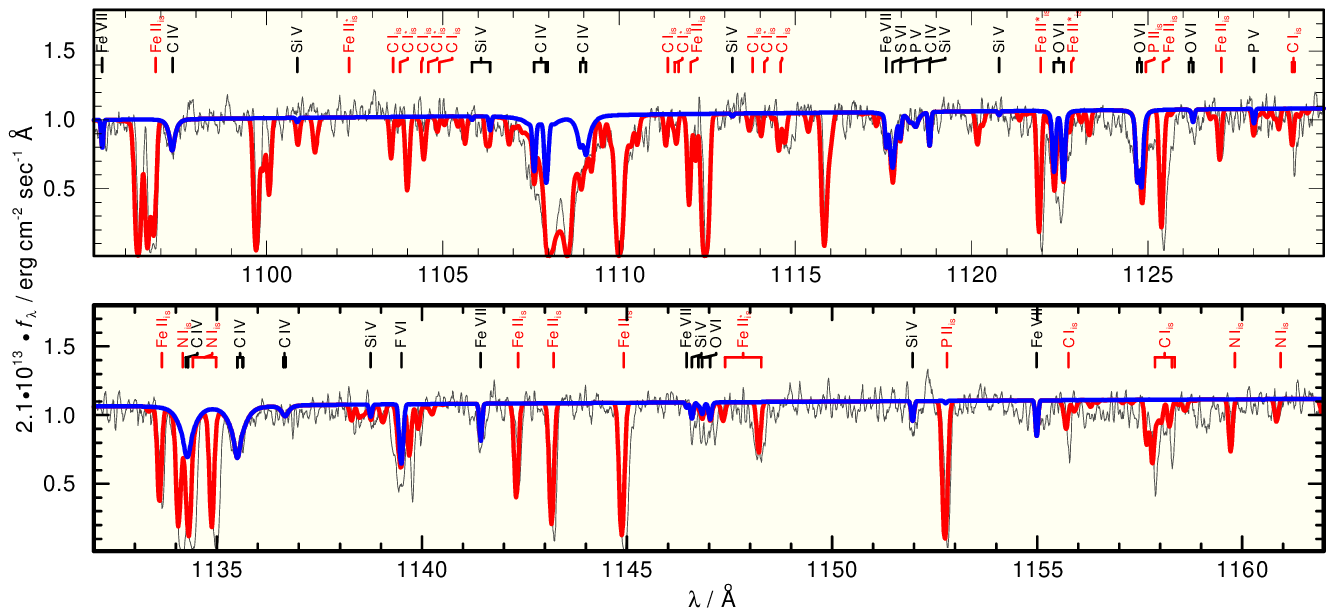}
\caption{Section of the \emph{FUSE} observations of A\,43 compared with a \emph{TMAP} (blue) and \emph{TMAP}/OWENS model (red).}
\label{FUSE}
\end{figure}

\section{Preliminary results and discussion}
The analysis yields slightly different values $T_{\mathrm{eff}}=105\,$kK$\,\pm10\,$kK, $\log g=5.6\pm0.3$, and different abundances (Table\,\ref{results}). Within the error limits these parameters agree with those of the CSPN of NGC\,7094. A mass of 0.53\,M$_\odot$ (evolutionary tracks, \citealt{MillerBertolami, MillerBertolami-Althaus}) and a distance of 2.2\,kpc were determined.
\begin{table}[!ht]
\caption{Abundances of the CSPN of A\,43, X is given in mass fractions, $\left[\mathrm{X}\right]$ denotes log $\frac{\mathrm{abundance}}{\mathrm{solar  \,\,abundance}}$. The typical error range is $\approx$0.3\,dex.}
\label{results}
\smallskip
\begin{center}
{\scriptsize
\begin{tabular}{rccccccccc}
\tableline
\noalign{\smallskip}
&H&He&C&N&O&F&Si&P&S\\
\noalign{\smallskip}
\tableline
\noalign{\smallskip}
X&0.24&0.56&0.19&2.4E-4&1.8E-3&2.8E-6&1.8E-3&1.3E-6&1.2E-3\\
$\left[\mathrm{X}\right]$&$-$0.483&$-$0.350&1.909&$-$0.491&$-$0.516&0.694&$-$0.560&$-$0.593&$-$0.378\\
\noalign{\smallskip}
\tableline
\end{tabular}
}
\end{center}
\end{table}

\begin{minipage}{7.9cm}
\setlength{\textwidth}{7.9cm}
\setlength{\textheight}{4.0cm}
\hspace{-1.0cm}
\plotone[width=\textwidth]{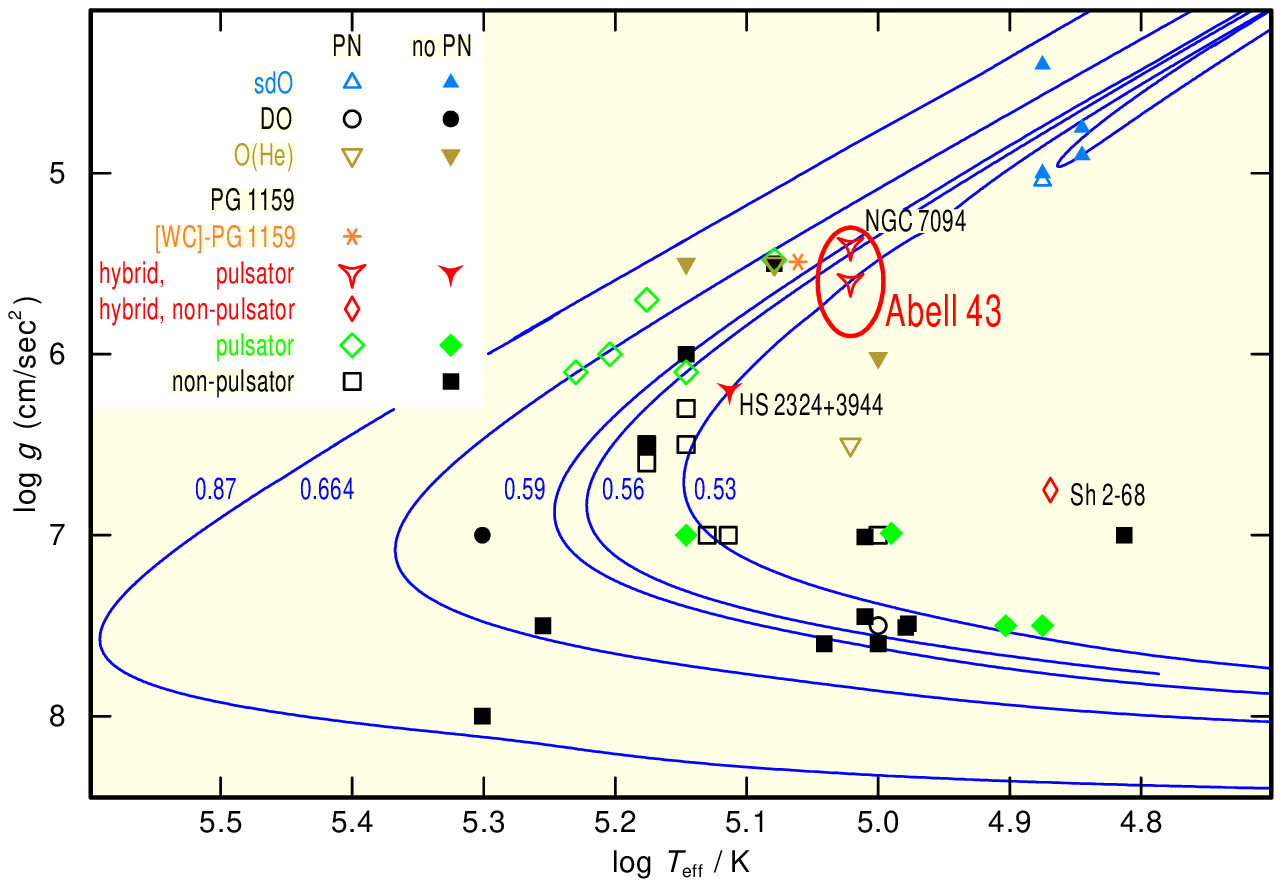}
\end{minipage}
\hbox{}\hspace{5mm}
\begin{minipage}{3.8cm}
\vspace{2.5cm}
\setlength{\textwidth}{3.8cm}
\setlength{\textheight}{2.7cm}
Figure 3. \,Location of A\,43 and related objects in the log $T_{\mathrm{eff}}$-$\log g$ plane.
\end{minipage}

\subsection{The German Astrophysical Virtual Observatory}
The German Astrophysical Virtual Observatory (GAVO) aims to make astronomical data accessible. It is funded by the Federal Ministry for Education and Research (BMBF) and operates within the International Virtual Observatory Alliance (IVOA). In the framework of a GAVO project the model-atmosphere code \emph{TMAP} was made accessible in different ways via \textit{TheoSSA} (Theoretical Simple Spectra Access). It provides:
\newline $\bullet$ SEDs (\textit{TheoSSA}, http://vo.ari.uni-tuebingen.de/ssatr-0.01/TrSpectra.jsp?)
\newline $\bullet$ Simulation Software (\textit{TMAW}, http://astro.uni-tuebingen.de/\raisebox{.2em}{\tiny $\sim $}TMAW/TMAW.shtml)
\newline $\bullet$ Atomic Data (\textit{TMAD}, http://astro.uni-tuebingen.de/\raisebox{.2em}{\tiny $\sim $}TMAD/TMAD.html)
\newline
It is controlled via a web interface where the fundamental parameters like $T_\mathrm{eff}$ or $\log g$ are entered. As a result a table of already available spectral energy distributions (SEDs) within a parameter range is given which can be downloaded directly. If a requested SED is not available, it can be calculated via \textit{TMAW}, the web interface of \emph{TMAP}. The database is growing in time because newly calculated SEDs are automatically ingested. In this way, the \emph{TMAP} code can easily be used for spectral analysis by everybody. This simplifies getting reliable fluxes of central stars that are e.g. necessary to model planetary nebulae properly.

\acknowledgements TR is supported by the German Aerospace Center (DLR) under grant
05\,OR\,0806. ER is supported by the Deutsche Forschungsgemeinschaft (DFG) under grant
WE\,1312/41-1. This work has been done using the profile fitting procedure OWENS.f developed by M. Lemoine and the \emph{FUSE} French Team.

\bibliography{Ringat}

\begin{thebibliography}{}

\bibitem[\protect\astroncite{{Fitzpatrick}}{1999}]{Fitzpatrick}
{Fitzpatrick}, E.~L. 1999,
\newblock {PASP}, {111}, 63

\bibitem[\protect\astroncite{{Miller Bertolami} \&
  {Althaus}}{2006}]{MillerBertolami}
{Miller Bertolami}, M., \& {Althaus}, L. 2006,
\newblock {A\&A}, {454}, 845

\bibitem[\protect\astroncite{{Miller Bertolami} \&
  {Althaus}}{2007}]{MillerBertolami-Althaus}
{Miller Bertolami}, M., \& {Althaus}, L. 2007,
\newblock {A\&A}, {470}, 675

\bibitem[\protect\astroncite{{Rauch} \& {Deetjen}}{2003}]{RauchDeetjen}
{Rauch}, T., \& {Deetjen}, J.~L. 2003,
\newblock in {Stellar Atmosphere Modeling}, edited by {I.~Hubeny, D.~Mihalas, \& K.~Werner}, vol. 288 of {ASPCS}, 103

\bibitem[\protect\astroncite{{Werner} et~al.}{2003}]{Werner2003}
{Werner}, K., {Deetjen}, J.~L., {Dreizler}, S., {Nagel}, T., {Rauch}, T., \&
  {Schuh}, S.~L. 2003, in {Stellar Atmosphere Modeling}, edited by {I.~Hubeny, D.~Mihalas, \& K.~Werner}, vol. 288 of {ASPCS}, 31

\bibitem[\protect\astroncite{{Werner} \& {Dreizler}}{1999}]{WernerDreizler}
{Werner}, K., \& {Dreizler}, S. 1999,
\newblock {Journal of Computational and Applied Mathematics}, {109}, 65

\bibitem[\protect\astroncite{{Werner} \& {Herwig}}{2006}]{WernerHerwig}
{Werner}, K., \& {Herwig}, F. 2006,
\newblock {PASP}, {118}, 183

\end{thebibliography}
\bibliographystyle{asp2010}

\end{document}